\documentstyle[12pt,twoside,fleqn,espcrc1,psfig]{article}

\newcommand{\AmS}{{\protect\the\textfont2
  A\kern-.1667em\lower.5ex\hbox{M}\kern-.125emS}}

\title{Kaon Electromagnetic Production on Nuclei}

\author{C. Bennhold\address{Department of Physics, George Washington
        University, Washington, D.C. 20052, USA},
        F.X. Lee\address{Department of Physics, University of Colorado, 
        Boulder, Colorado 80309-0446, USA},
        T. Mart\address{Jurusan Fisika, FMIPA, 
        Universitas Indonesia, Depok 16424, Indonesia},
        L.E. Wright\address{Department of Physics, Ohio University,
        Athens, Ohio 45701, USA}}

\begin{document}
\maketitle

\begin{abstract}
The formation and excitation of hypernuclei through kaon photoproduction is reviewed.
Basic features of the production process are emphasized.
  The possibility of extracting new information on
hypernuclear structure and on the wave function of the bound $\Lambda$ 
is discussed.  New results are presented for the quasifree production
process $A(\gamma, K \Lambda)B$.  Observables of this reaction are
shown to be sensitive to the $\Lambda$-nucleus final state interaction.
\end{abstract}

\section{HYPERNUCLEAR EXCITATION}

With the commissioning of Jefferson Lab and other continuous beam electron
linacs with sufficient energy and intensity the exploration of hypernuclear
structure through the electromagnetic probes is becoming a reality.
In contrast to the hadronic reactions $(K^-, \pi^-)$ and $(\pi^+,K^+)$, the 
$(\gamma, K^+)$ process uses rather weakly interacting probes, the photon
and the $K^+$, with its mean free path of 5-7 fm in the nuclear medium,
allowing the process to occur deep in the nuclear interior.  In comparison,
the $K^-$ and the $\pi^{\pm}$ are both strongly absorbed, thereby confining
the reaction to the nuclear periphery.   
Due to the mass difference in the incoming kaon and outgoing pion,
the $(K^-,\pi^-)$ reaction allows for recoilless $\Lambda$ production
in the nucleus, leading to high counting rates.  Kaon photoproduction,
on the other hand, involves high momentum transfers due to the large
production of the rest mass which will therefore project out high
momentum components of the nuclear wave functions.
The subject of exciting discrete hypernuclear states through
kaon photoproduction was studied extensively about 8-10 years 
ago\cite{bennhold89,cotanch86,rosenthal88,cohen89} but has been
mostly dormant for the last several years.  However, a number of
planned and approved experiments to take place
within the next few years has the potential to revive interest
in this field.

\subsection{Matrix elements for the process
$\gamma + {\rm A} \rightarrow K^+ + 
~_\Lambda{\rm B}$ }

As shown in detail in ref.\cite{bennhold89}, the differential
cross section for the reaction $\gamma + {\rm A} \rightarrow K^+ + 
~_\Lambda{\rm B}$ in the center of momentum (cm) frame can be written as
\begin{eqnarray}
\frac{d\sigma}{d\Omega^{\rm cm}_{K}} &=& \frac{1}{16\pi^2} \frac{q^{\rm cm}}{
k^{\rm cm}} \frac{m_im_f}{W^2} \frac{F_{\rm cm}}{2(2J_i+1)} \sum_{M_i,M_f,
\lambda} |\langle J_fM_f,T_fN_f;K^+|T|J_iM_i,T_iN_i;\gamma \rangle |^2 ~,
\end{eqnarray}
where we average over the initial spin projection $M_i$ as well as the photon
polarization $\lambda$ and sum over the final spin projection $M_f$. The 
four-momenta of the photon and kaon are denoted by $(E_{\gamma},\bf k)$
and $(E_K,\bf q)$, the total spin and isospin are $J_i,J_f$ and $T_i,T_f$,
respectively, along with their projections $M_i,M_f$ and $N_i,N_f$. The
usual correction factor compensating for the lack of translational invariance
of the shell model is given by $F_{\rm cm}={\rm exp}[b^2({\bf k} - {\bf q})^2
/2A]$, $b$ being the harmonic oscillator parameter for the nucleus under
study and $A$ the nuclear mass number. The masses of the initial- and 
final-state nuclei are $m_i$ and $m_f$, and $W$ is the total energy in 
the cm system.

Taking the nuclear kaon production amplitude to be a one-body operator
the nuclear matrix element is given by

\begin{eqnarray}
\langle J_fM_f,T_fN_f;K^+|T|J_iM_i,T_iN_i;\gamma \rangle &=& \sum_{\alpha ,
\alpha '} \langle J_fM_f,T_fN_f;K^+|C_{\alpha '}^\dagger C_{\alpha}
|J_iM_i,T_iN_i;\gamma \rangle~ \nonumber\\
&& \times ~\langle \alpha ' ;K^+|t|\alpha ;\gamma
\rangle ~.
\label{eq:basic}
\end{eqnarray}
In Eq.~(\ref{eq:basic}) the many-body nuclear structure aspects are already
separated from the photoproduction mechanism but in principle the sum
extends over a complete set of single-particle states $\alpha$ and
$\alpha '$. The nuclear structure information involved in one-body 
processes is usually contained in the double-reduced density matrix elements
(RDME),
\begin{eqnarray}
\Psi_{J;T}(a',a) &=& {\hat J}^{-1}{\hat T}^{-1}
\langle J_f,T_f || [C_{\alpha '} \otimes C_{\alpha}] ||J_i,T_i \rangle ~.
\end {eqnarray}

All the dynamics of the photoproduction process is contained in the
single-particle matrix element $\langle \alpha ' ;K^+|t|\alpha ;\gamma
\rangle $ which in general involves a nonlocal 
operator. In momentum space this matrix element has the form
\begin{eqnarray}
\langle \alpha ' ;K^+|t|\alpha ;\gamma \rangle &=& \int d\!~^3p~ d\!~^3q'~
{\psi}^{*}_{\alpha '}({\bf p'}) \phi^{*(-)}_{K} ({\bf q},{\bf q'})
t_\gamma \psi_{\alpha}({\bf p}) ~,
\label{eq:sixdim}
\end{eqnarray}
where $\bf p' = \bf p + \bf k - \bf q$, and $\psi$ is the single-particle
wave function of the proton in the initial and the $\Lambda$ in the final state.
The wave function with the
appropriate boundary conditions for the outgoing kaon of three-momentum
$\bf q$, distorted by its interaction with the residual hypernucleus
through an optical potential, is denoted by $\phi^{*(-)}_K (\bf q,\bf q')$.
This wave function is generated by solving the Klein-Gordon
equation using a simple $t \rho$ optical potential with the $K^+ N$ phase shifts 
of ref.\cite{martin75}.

\subsection{Basic features of the coherent kaon production process}

\begin{figure}
\centerline{\psfig{file=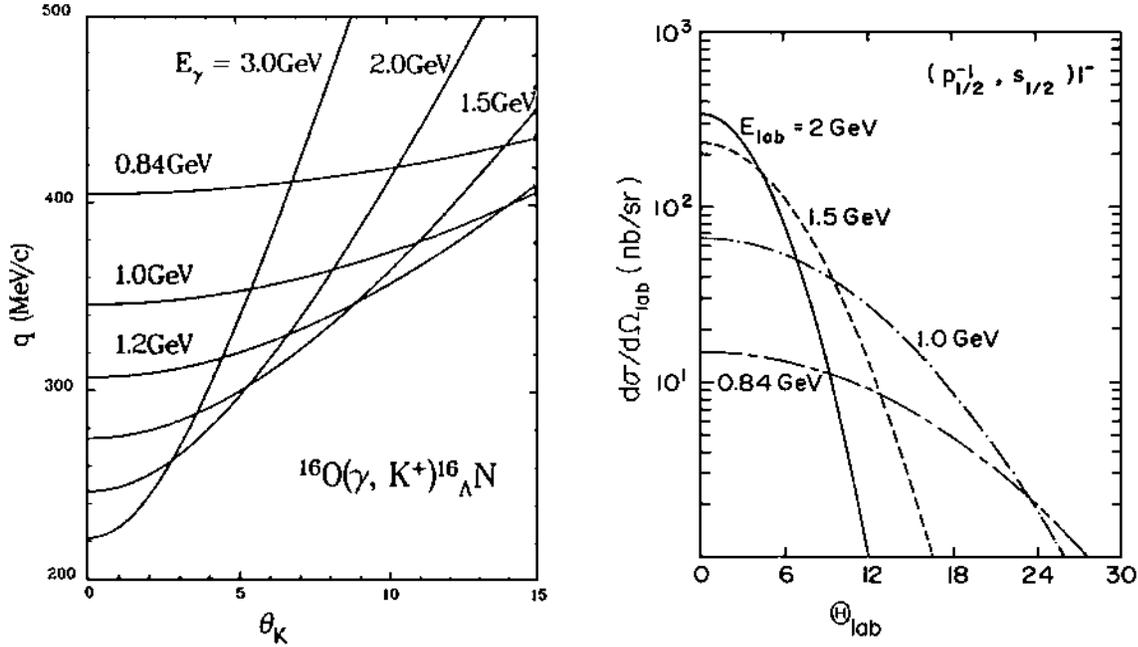,width=6in}}
\caption{Kinematic features of the $(\gamma, K^+)$ process. The left side
shows the momentum transfer in the lab system for several photon
energies $E_{\gamma}$.  The right side shows angular distributions for
the $1^-$ member of the ground state doublet in
$^{16}O(\gamma,K^+)^{16}_{\Lambda}N$ for various photon lab energies.}
\label{h97nucf1}
\end{figure}

Fig.~\ref{h97nucf1} compares the momentum transfer behavior 
with the magnitude of the
differential cross section for the reaction
$^{16}O(\gamma,K^+)^{16}_{\Lambda}N$.
At $0^o$ kaon lab angle the momentum transfer to
the final hypernuclear system decreases as the photon lab energy
increases.  This leads to a differential cross section at $\Theta_K=0^o$
which increases as $E_{\gamma}$ increases, from around 15 nb/sr at 0.84 GeV
to 330 nb/sr at 2 GeV for the particular transition shown.  However, the momentum transfer increases more rapidly for non-zero kaon angles at higher photon energies.
Thus, the angular distributions become more forward peaked and fall off
more rapidly.  The energy chosen for an experiment therefore depends
on the desired result: if the goal is to perform hypernuclear
spectroscopy choosing a higher photon energy around 2 GeV,
while detecting the $K^+$
under $0^o$ would be advantageous. If, on the other hand, one likes to
extract dynamical information by mapping our transition densities via
measuring angular distributions, photon energies of 1.0 to 1.2 GeV would
be preferable.

With the exception of ref.\cite{rosenthal88,motoba94} all calculations
up to now have been performed in pure particle-hole configurations.
These predictions may be reliable  where
the proton pick-up strength is not highly fragmented 
for stretched spin-flip transitions with maximum $J=l_N + l_{\Lambda} +1$.
These transitions are usually dominated by the Kroll-Ruderman 
{\boldmath $\sigma\cdot\epsilon$} operator 
and tend to have the largest cross sections.
In the case shown in Fig.~\ref{h97nucf1} we describe $^{16}_{\Lambda}N$ as a
pure $p_{1/2}$ proton hole coupled to an s-shell $\Lambda$, coupling
to a $0^-$ and $1^-$ ground state transition. The degeneracy between these
two states would be removed by including the  $\Lambda N$ interaction.
For a closed shell target nucleus in a pure particle-hole basis, the
RDME simply reduce to
$\Psi_{J;T}(a',a) = \delta_{ab} \delta_{a'b'}$.  While the stretched transitions
are most likely the first ones to be measured, eventually one would like
to use the $(\gamma, K^+)$ reaction to extract hypernuclear structure
information from cases where configuration mixing is important.  As discussed in
ref.\cite{millener90}, the reaction
$^{9}Be(\gamma,K^+)^{9}_{\Lambda}Li$ may provide a good testing ground
for resolving members of the $s_{\Lambda}$ doublet of $3/2^+$ and $5/2^+$,
coupling the s-shell $\Lambda$ to the $2^+$ core of ${^8}Li$.
This doublet is split by 0.51 MeV according to 
the "standard" $\Lambda N$ interaction of ref.\cite{millener85}.  The predicted
$\Delta S$ = 0 RDME are large for the lower member of the doublet but
small for the upper member while there is a large $\Delta J$ = 2 ($\Delta S$ = 1)
RDME for the upper member; thus, the two transitions would produce very different
angular distributions.  Similar information may be extracted from the reaction
$^{13}C(\gamma,K^+)^{13}_{\Lambda}B$. 

\subsection{The role of the elementary operator}

Over the last several years considerable effort has been spent to
develop models for the electromagnetic production of kaons from
nucleons at photon energies 
below 2 GeV\cite{adelseck85,david96,williams92,mart95}. Most of these 
analyses have
focused on the two processes $\gamma p\rightarrow K^{+}\Lambda $
and $\gamma p\rightarrow K^{+}\Sigma ^0 $, using real or virtual
photons, since almost all of the few available photokaon data have
been taken for these two reactions. Due to the limited set of data
the various models permit only qualitative
conclusions but do not yet allow the extraction of precise coupling
constants and resonance parameters.

\begin{figure}[thb]
\centerline{\psfig{file=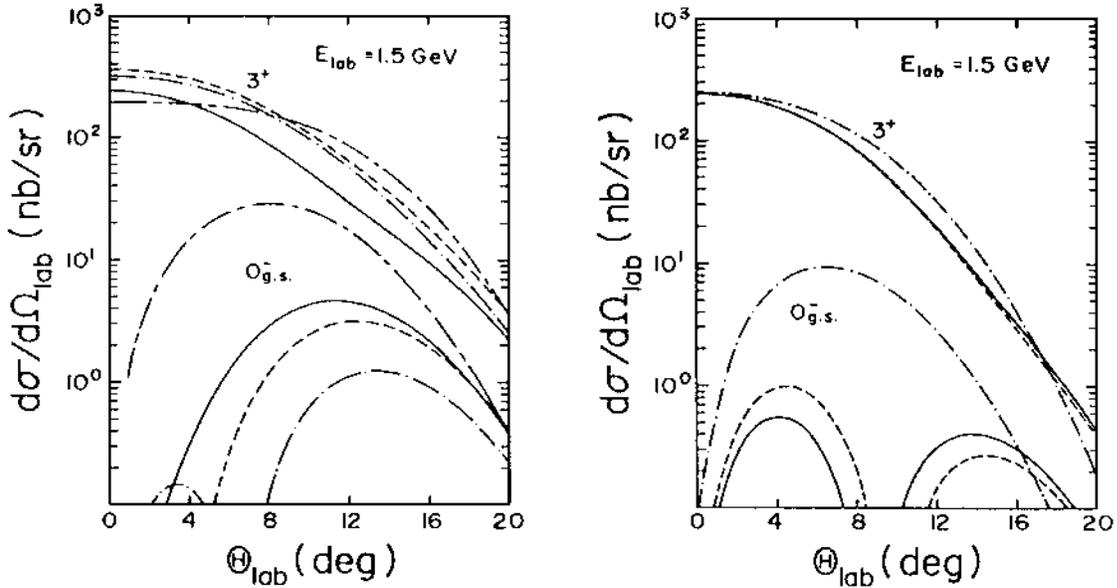,width=6in}}
\caption{Sensitivity of the hypernuclear cross section 
to the elementary operator.
The left side compares angular distributions for
$^{16}O(\gamma,K^+)^{16}_{\Lambda}N$ using different elementary amplitudes 
(see ref.\protect\cite{bennhold89} for details), 
while the right side compares the full operator (dashed curve) 
to the nonrelativistic reduction (solid curve) and the
{\boldmath $\sigma\cdot\epsilon$} term only (dash-dotted curve).}
\label{h97nucf2}
\end{figure}

Fig.~\ref{h97nucf2} illustrates the sensitivity of the hypernuclear 
formation cross sections
to different elementary amplitudes for the stretched $3^+$ excited state
and the $0^-$ ground state transition.  
All the amplitudes used give an equally
good fit to the available cross section data on the nucleon.
The results for the $3^+$ state are similar
in shape and differ by at most 50\% in magnitude while different operators
can lead to angular distributions differing by more than an order of magnitude.
This can be traced to the dominance of the Kroll-Ruderman term in the $3^+$
transition which is shown to deviate from the full operator 
only slightly in the right panel of Fig.~\ref{h97nucf2}.  
The $0^-$ state, on the other hand,
is clearly very sensitive to details of the non spin-flip 
pieces of the operator.
Retaining the {\boldmath $\sigma\cdot\epsilon$} term only 
overestimates the cross section
by more than an order of magnitude and misses the minimum
from the form factor.  Clearly, the elementary process has to be
measured more precisely before reliable predictions can be made.

The stretched transitions such as the $3^+$ state discussed above
can be used to extract the bound $\Lambda$ wave function.  If a transition
is dominated by the Kroll-Ruderman term, the operator becomes local
and can be factored out of the single-particle matrix element of Eq. 4.
Furthermore, neglecting kaon distortion which reduces cross sections only by
about 10-20\% for p-shell nuclei reduces the matrix element to

\begin{eqnarray}
\langle \alpha ' ;K^+|t|\alpha ;\gamma \rangle & = & const. \int r^2 dr~ 
{\psi}^{*}_{\Lambda}(r)  \psi_{p}(r) j_{L}(Qr) ~.
\end{eqnarray}

Therefore, assuming one has good knowledge of the bound proton wave function
from $(e,e' p)$, measuring an angular distribution will map out the bound
$\Lambda$ wave function.  This feature is unique to the $(\gamma, K^+)$ process
since distortion effects are minimal.

\section{EXCLUSIVE QUASIFREE KAON PHOTOPRODUCTION}

Due to the sizable momentum transfer to the hypernuclear system
the probability of forming such bound states is in fact rather small.
Ref.\cite{cotanch86} has estimated this formation probability
to be around 5-10\% of the total $(\gamma, K^+)$ strength
on nuclear targets, thus most of the kaon production events
will come from quasifree production.  Here
we present the first theoretical predictions for quasifree
kaon photoproduction from nuclei, $A(\gamma, K Y)B$, where the
kaon can be a $K^+$ or $K^0$, and the hyperon can be either
a $\Lambda$ or a $\Sigma$.  Thus, if this process is measured
exclusively six possible reaction channels can be explored.
The predictions are made in a DWIA framework that has been
successfully applied in our previous work on quasifree pion
photo- and electroproduction\cite{lee93} and eta photoproduction
on nuclei\cite{lee96}.  This reaction allows for the study of the reaction
process in the nuclear medium as well as final state interaction effects
without being obscured by the details of the nuclear transitions
as discussed above.  This is due mainly to the quasifree nature
of the reaction which permits the kinematic flexibility to have small
momentum transfers.  The key ingredients are: a) the single-particle
wave function and spectroscopic factor, usually taken from electron scattering,
b) the elementary kaon photoproduction amplitude, obtained from measuring the
the free processes, c) the distorted kaon wave function which can be taken from
kaon elastic scattering in case of the $K^+$, and finally, d) the hyperon-nucleus
final-state interaction.  It is mainly to study the last ingredient
that we investigate this process which is already part of an approved
TJNAF experiment in Hall B.

\subsection{Matrix elements and observables}

The coordinate system is defined such that
the z-axis is along the photon direction ${\bf k}$,
and the y-axis is along ${\bf \hat{k}} \times {\bf \hat{q}}$ with
the azimuthal angle of the kaon chosen as $\phi_K = 0$.
The kinematics of the reaction are determined by
${\bf k}={\bf q}+{\bf p}+{\bf Q}$ and
$E_\gamma +M_i = E_K +E_Y +M_f + T_Q$.
Here ${\bf Q}$ and $T_Q=\frac{Q^2}{2M_f}$ are the
missing momentum and missing kinetic energy in the reaction,
respectively, and $|{\bf k}|=E_\gamma$ for real photons.
The binding energies are included in the masses $M_i$ and $M_f$.
In the impulse approximation,
the reaction is assumed to take place on a single bound nucleon whose
energy and momentum are given by
${\bf p}_i =-{\bf Q}$ and $E_i=E_K +E_Y -E_\gamma$.
The struck nucleon is in general off its mass shell.

In contrast to hypernuclear production discussed above,
the reaction is {\em quasifree}, meaning that the magnitude of
${\bf Q}$ has a wide range, including zero.
Since the reaction amplitude is proportional to the Fourier transform 
of the bound state single particle wavefunction, it falls off quickly 
as the momentum transfer increases. 
Thus, we will restrict ourselves to the low $Q$ region ($<$ 500 MeV/c)
where the nuclear recoil effects can be safely neglected 
for nuclei of $A > 6$.

The differential cross section can be written as 
\begin{equation}
\frac{d^3 \sigma}{dE_K\,d\Omega_K\,d\Omega_Y}=
{C \over 2(2J_i+1)}\;
\sum_{\alpha,\lambda,m_s}\frac{S_\alpha}{2(2j+1)}
|T(\alpha,\lambda,m_s)|^2,
\label{coin}
\end{equation}
where the kinematical factor is given by
\begin{equation}
C=
{ M_f m_Y\, |{\bf q}|\,|{\bf p}| \over
4(2\pi)^5 
|E_Y +M_f+T_Q -E_Y \,{\bf p}\cdot({\bf k}-{\bf q})/p^2| }.
\end{equation}
The single particle matrix element is given by
\begin{equation}
T(\alpha,\lambda,m_s) = \int d^3 r\,
\Psi^{(+)}_{m_s}({\bf r},-{\bf p})\;
\phi^{(+)}_K({\bf r},-{\bf q})\;
t_{\gamma K}(\lambda, {\bf k}, {\bf p}_i ,{\bf q}, {\bf p})\;
\Psi_{\alpha}({\bf r})\;
e^{i{\bf k}\cdot{\bf r}}.  
\label{3d}
\end{equation}
In the above equations, $J_i$ is the target spin, 
$\alpha=\{nljm\}$ represents the single particle states,
$S_\alpha$ is called the spectroscopic factor whose value is
taken from electron scattering,
$\lambda$ is the photon polarization, $m_s$ is
the spin projection of the outgoing nucleon, $\Psi^{(+)}_{m_s}$
and $\phi^{(+)}_K$ are the distorted wavefunctions with outgoing
boundary conditions, $\Psi_{\alpha}$ is the bound nucleon wavefunction,
and $t_{\gamma K}$ is the kaon photoproduction one-body operator.

We also compute the photon asymmetry, defined by
\begin{equation}
\Sigma=
{ 
{d^3 \sigma}_{\perp}-{d^3 \sigma}_{\parallel}
\over
{d^3 \sigma}_{\perp}+{d^3 \sigma}_{\parallel}
}
\end{equation}
where $\perp$ and $\parallel$ denote the perpendicular and parallel
photon polarizations relative to the production plane (x-z plane),
and the hyperon recoil polarization (also referred to as analyzing power) 
defined by
\begin{equation}
P=
{ 
{d^3 \sigma}_{\uparrow}-{d^3 \sigma}_{\downarrow}
\over
{d^3 \sigma}_{\uparrow}+{d^3 \sigma}_{\downarrow}
}
\end{equation}
where $\uparrow$ and $\downarrow$ denote the polarizations 
of the outgoing hyperon relative to the y-axis.
We have used the short-hand notation
$d^3 \sigma \equiv d^3 \sigma/dE_K\,d\Omega_Kd\Omega_Y$ with
appropriate sums over spin labels implied.
Note that $P$ is obtained for free since the produced hyperon is 
self-analyzing, 
while the measurement of $\Sigma$ requires a linearly polarized photon beam.

For the bound nucleon, we use harmonic oscillator wavefunctions
which are sufficient in the quasifree region we are interested in.
For the outgoing kaon the same distorted wave function is used as
in hypernuclear production discussed above.

\subsection{The hyperon-nucleus optical potentials}
\begin{figure}[p]
\centerline{\psfig{file=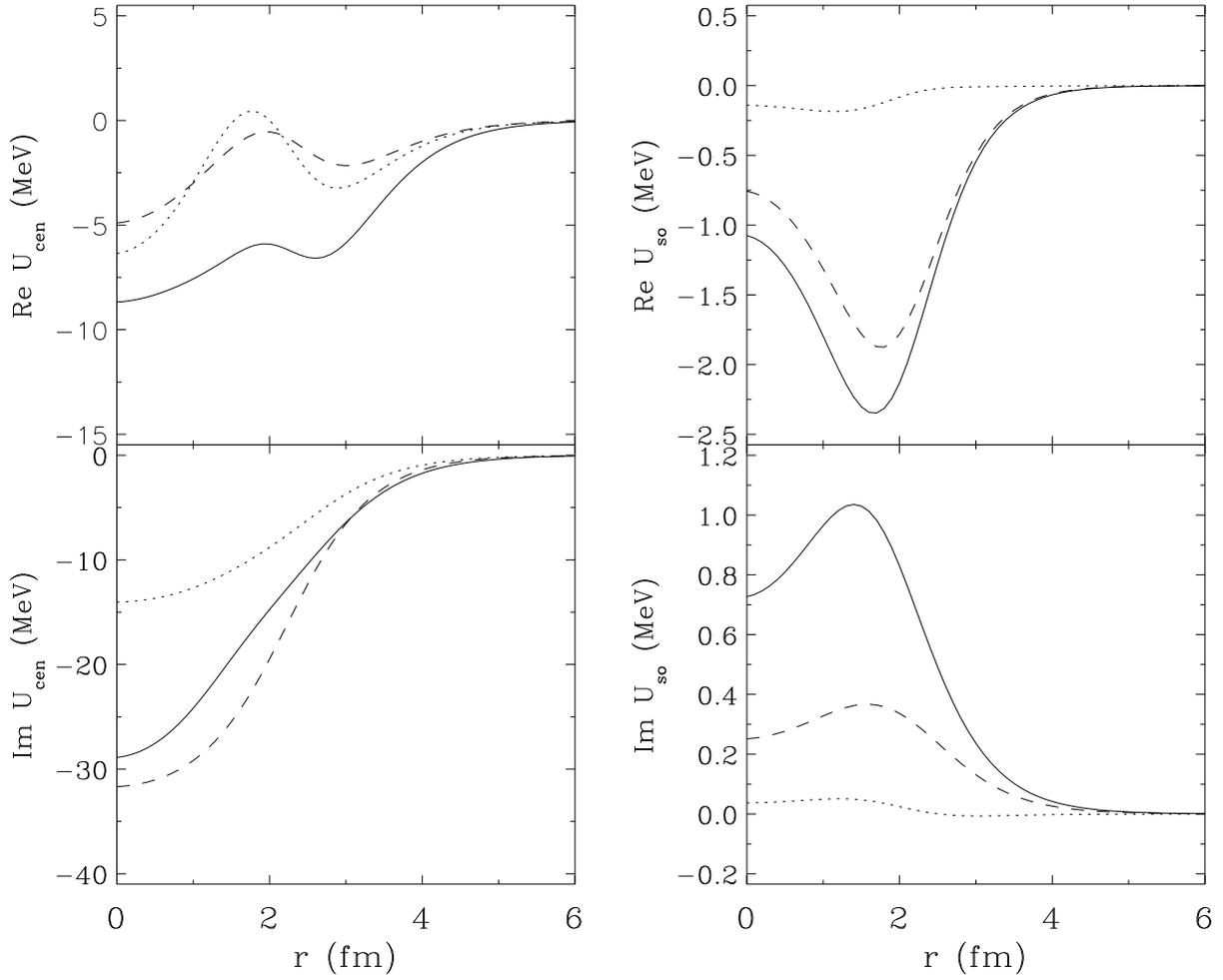,width=6in}}
\caption{Hyperon optical potentials for $^{12}C$ at 200 MeV kinetic
energy.
The dashed line shows the $\Lambda$ potential, while the dotted line 
depicts the $\Sigma^0$ potential.
The proton potential (solid line) is shown for comparison.}
\label{h97nucf3}
\end{figure}

Very few optical potentials have been constructed for the
$\Lambda$ and $\Sigma$, mostly due to lack of data.
Here, we employ the global optical model of ref.\cite{cooper94}.
It is based on a global nucleon-nucleus
Dirac optical potential fit\cite{cooper93}.
The parameters of the potential are motivated by the
constituent quark model and adjusted to fit the hypernuclear
binding energy data
We use its nonrelativistic equivalent version which has a
central part and a spin-orbit part:
$U(r)=U_{cen}(r)+U_{so}(r)\; {\bf \sigma \cdot l}$.
Note that the spin-orbit part is multiplied by a factor that depends on
the partial wave under consideration.
Fig.~\ref{h97nucf3} shows the real and imaginary parts of both
$U_{cen}(r)$ and $U_{so}(r)$ on $^{12}C$ at 200 MeV kinetic energy
for the $\Lambda$ and the $\Sigma^0$. For comparison, they are also shown for proton.
The real parts of the central potential are clearly smaller than the
proton potential by around a factor of two, reflecting the fact
that Lambdas have a smaller binding energy in hypernuclei
while Sigmas appear not to be bound at all.  The imaginary part
of the $\Sigma$'s central potential is similar in magnitude to that of the nucleon,
due to the large $\Sigma N \rightarrow \Lambda N$ conversion width.
The very small spin-orbit potential of the $\Lambda$ is a reflection
of the $\Lambda N$ spin-orbit force which is known to be small.

\subsection{Results and discussion}

\begin{figure}[p]
\centerline{\psfig{file=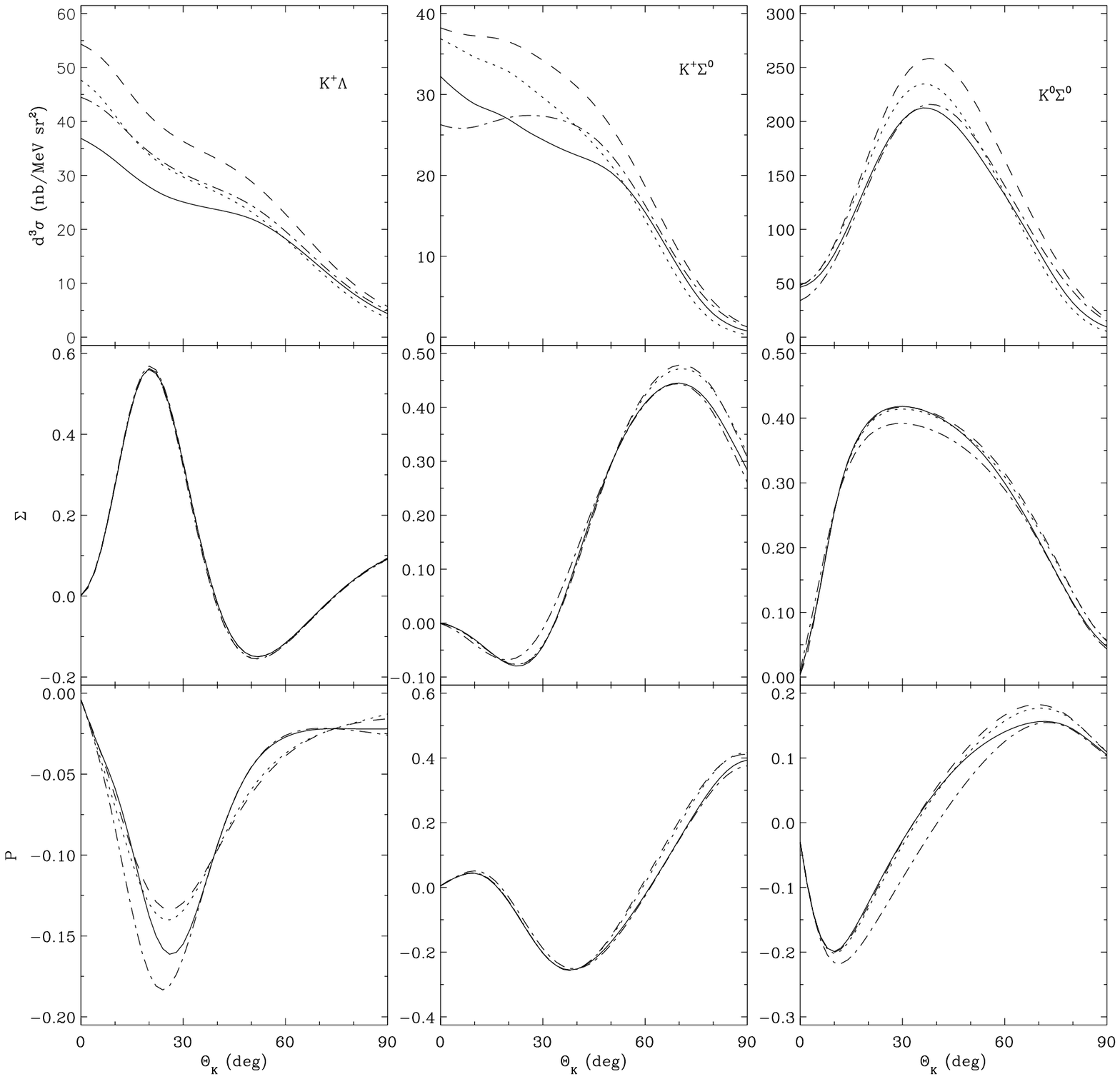,width=5.9in}}
\caption{Results for the reaction $^{12}C(\gamma,KY)B_{g.s.}$ 
at $E_\gamma$=1.4 GeV and Q=120 MeV/c under quasifree kinematics.
Three of the six possible channels are shown in the three columns,
using the model of Ref.~\protect\cite{williams92} for the elementary
process.
The four curves correspond to calculations in PWIA (dashed),
in DWIA with kaon only distorted (dotted),
with hyperon only distorted (dash-dotted), and with both distorted (solid).}
\label{h97nucf4}
\end{figure}

Our results are shown in Fig.~\ref{h97nucf4} for the reaction
$^{12}C(\gamma,KY) ^{11}B_{g.s.}$
at $E_\gamma$=1.4 GeV and Q=120 MeV/c under quasifree kinematics.
Quasifree kinematics is similar to two-body kinematics in free space,
except that the nucleon is bound with finite momentum Q.

As the kaon angle increases,
the kaon energy decreases, the hyperon energy increases,
the hyperon angle increases to a certain value, then decreases.
In this particular case, the kaon and hyperon energies reach around 500 to 700 MeV,
and the hyperon angle reaches up to  45 degrees, depending on the channel.
Fig.~\ref{h97nucf4} shows the differential cross sections as well as two polarization observables,
comparing PWIA calculations with results that include hyperon and kaon final state
interaction (FSI).  As in the case of hypernuclear production discussed above
kaon distortion reduces the cross sections by about 10-20\% but has little
effect on the polarization observables.  In the case of $K^0 \Sigma^0$ production
we have used the same optical potential as for the $K^+$ since little is known
about the $K^0$ nucleus interaction.  In principle, such information can be obtained
by measuring kaon charge exchange on nuclei.  Including the hyperon FSI
reduces the angular distributions  by up to 30\% at forward angles.  Again,
with the exception of the recoil polarization in $K^+ \Lambda$ production, the
polarization observables are barely affected by the inclusion of FSI.
 This situation is similar to
our previous findings in quasifree pion and eta photoproduction\cite{lee93,lee96}.
It opens the possibility to use the polarization observables as a way to
study modifications of the basic production process in the nuclear medium.
The magnitudes of the $\Sigma$ and $P$ observables is sizeable and should
 be measurable.  However, they are very sensitive
to the elementary operator used, thus, given our limited knowledge of the
elementary process, the polarization observables shown here should be
treated with caution.  The first set of CLAS experiments with
new data for the kaon production reactions on the nucleon will allow
a precise determination of the elementary amplitude which in turn
can be used to reliably predict these observables.

\end{document}